# Dynamic Behavior Control of Induction Motor with STATCOM

Majid Dehghani
Dept. of Electrical Engineering
Amirkabir University of Technology
Tehran, Iran
majid1369@aut.ac.ir

Peyman Karimyan
Dept. of Electrical Engineering
Amirkabir University of Technology
Tehran, Iran
peyman.sena @gmail.com

Mehradad Abedi
Dept. of Electrical Engineering
Amirkabir University of Technology
Tehran, Iran
abedi@aut.ac.ir

Hadis Karimipour
School of Enginering
University of Guelph
Guelph, Canada
hkarimi@uoguelph.ca

*Abstract*— **STATCOMs is used widely in power systems these days. Traditionally, this converter was controlled using a double-loop control or Direct Output Voltage (DOV) controller. But DOV controller do not function properly during a three-phase fault and has a lot of overshoot. Also, the number of PI controllers used in double-loop control is high, which led to complexities when adjusting the coefficients. Therefore, in this paper, an improved DOV method is proposed which, in addition to a reduced number of PI controllers, has a higher speed, lower overshoots and a higher stability in a wider range. By validating the proposed DOV method for controlling the STATCOMs, it has been attempted to improve the dynamical behaviors of induction motor using Matlab/Simulink, and the results indicate a better performance of the proposed method as compared to the other methods.**

I. INTRODUCTION

Nonlinear loads used in the industry exert a negative impact on the quality of the power delivered to the consumer. Voltage fluctuations are one of those phenomena that have an adverse effect on the system's sensitive equipment and cause problems in the system[2]. In the past, reactive power compensators, such as TCRs or capacitor banks, were used to solve this problem [3]. With the development of electronic devices, a new generation of static compensators (STATCOMs) were introduced that utilize a PWM voltage source converter and connect in parallel to the transmission or distribution system [4]. A STATCOM, in contrast to an SVC, has a better performance in reactive power compensation [5] and is a better voltage stabilizer at PCC point[6]. Accordingly, throughout recent years, the use of STATCOMs has been of interest to researchers to improve the power quality of electrical power systems. A STATCOM, similar to an SVC, can operate at higher speeds and with better dynamic characteristics, and, unlike the SVC, it does not depend on the network voltage. This is important when a fast dynamic response is required or the network voltage is low; in this case, the STATCOM has a better performance than the SVC. The functioning of the STATCOM is not dependent only on topology, but also on the way it is controlled [3]. Researchers focus on topology and control strategy in order to improve the performance of the STATCOM. These strategies should be used for bus voltage regulation, reactive power compensation and power factor correction.

Mehtn and Schauder [7]presented a Conventional Double-Loop Control Strategy. In this method, there are two loops; the external loop of the active and reactive current generates the reference current for maintaining the voltage at the PCC point, and the internal loop is used to control the inverter current[8].

Due to the use of both voltage and current sensors, four PI controllers are used in this method. Adjusting PI controller coefficients is a key factor in the performance of a STATCOM, on which a large number of studies have been conducted [9-14]. In [15], using the genetic algorithm, the neural network and the fuzzy-neural system, the PI controller coefficients for the STATCOM were investigated, and the results demonstrated that the performance of the fuzzy-neural controller was better than in other methods. In [16], a DSTATCOM was first modeled, and then a fuzzy system was employed to determine PI controller coefficients that improved the damping of the power system with a new controller.

However, this control strategy has four PI controllers, which not only complicate the design, but also bring about difficulties when adjusting its control parameters. These parameters are adjusted empirically or through trial and error, which either is time-consuming or cannot improve stability over a wide range of different system conditions [1].

To solve this problem, CHEN and Hsu [17], based on the theory of instantaneous power balance, proposed a Direct Output Voltage Control (DOV) for the STATCOM. This new controller eliminates the active and reactive feedback loop, reduces the number of PI controllers and calculates the STATCOM output voltage using an algebraic algorithm based on the principle of maintaining the power balance [18]. In [19], the DOV controller is used to correct the power factor and to reduce line harmonics. In [20], a direct output voltage control strategy is proposed based on a multi-modulator controller and a neural network. In this method, a neural network is used to adjust the values of the PI controller parameters according to the optimal control rule. Simulation results showed that, in comparison with the traditional PI controller, the PI controller is able to withstand a change in voltage with a higher compensation accuracy.

One of the disadvantages of this method is its inappropriate performance during a three-phase fault, which results in a high overshoot and low response speed. In this paper, an improved DOV method is presented to overcome the above problems. This control strategy can not only increase the response speed of the controller, but also reduce unwanted overshoots in the system during a three-phase fault.

In the second part of this paper, STATCOM-based reactive power compensation has been briefly described, and its performance-related characteristics and dynamical equations have been investigated. In the third section, the double-loop controller and the DOV controller have first been explained, and then the controller has been described. In the fourth section, the impact of these controllers on system performance is investigated. Finally, a conclusion has been made in the fifth part.

## II. THE STATIC SYNCHRONOUS COMPENSATOR

Static synchronous compensator is one of the instruments of FACTS. As can be seen from Fig. 1, STATCOM is connected to the transmission line in a parallel pattern [14, 21]. Voltage source converter exchanges the reactive power with the line by changing DC voltage to a variable AC voltage in the output. The level and direction of reactive power which is exchanged between STATCOM and the transfer line are identified based on the relative disparity of the level of STATCOM output voltages [22].

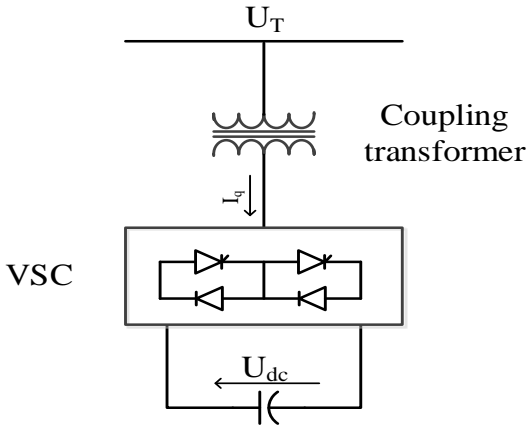

Fig. 1. STATCOM parallel connection to the network

### A. Dynamic characteristics of the STATCOM

A STATCOM model is shown in Fig 2. In this circuit, R represents the sum of the ohm losses of the transformer winding and the conductor losses in the converter, and L is the transformer leakage inductance [23]. Sum of the losses of the converter switching and the power loss in the capacitor is modeled with r as resistance, which is parallel to the DC link capacitor.

The voltages $e_a$, $e_b$ and $e_c$ are the phase voltages at the converter output. Inductors current of STATCOM are considered as state variable. By writing *KVL* equations at STATCOM output, following equation is obtained [24, 25]:

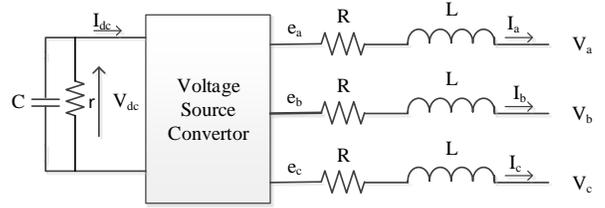

Fig. 2. The STATCOM equivalent model

$$L_S \frac{dI_{as}}{dt} = -R_s I_{as} + V_{as} - V_{al}$$

$$L_S \frac{dI_{bs}}{dt} = -R_s I_{bs} + V_{bs} - V_{bl} \quad (1)$$

$$L_S \frac{dI_{cs}}{dt} = -R_s I_{cs} + V_{cs} - V_{cl}$$

where $I_{as}$, $I_{bs}$ and $I_{cs}$ are three-phase currents, $V_{al}$, $V_{bl}$ and $V_{cl}$ are voltage vector of the PCC.

By utilizing Park's transformation for both sides of Eq. (1) and simplifying the relations, state equations for the phase current in the d-q coordinates are obtained as follows [26, 27]:

$$L_s \frac{dI_{ds}}{dt} = -R_s I_{ds} + wL_s I_{qs} + V_{ds} - V_{dl}$$
$$L_s \frac{dI_{qs}}{dt} = -R_s I_{qs} - wL_s I_{ds} + V_{qs} - V_{ql} \quad (2)$$

Where, $I_{ds}$, $I_{qs}$, $V_{ds}$, $V_{qs}$ are d- and q-axis STATCOM output currents and voltages respectively, $\omega$ is the synchronously rotating angle speed of the voltage vector of the PCC and $V_{dl}$, $V_{ql}$ are the load voltages.

## III. THE CONVENTIONAL DOUBLE-LOOP CONTROL STRATEGY

In the conventional control strategy, the typical PI controller is used to exchange active and reactive powers between the STATCOM and the network. The block of the conventional controller diagram is shown in Fig. 3.

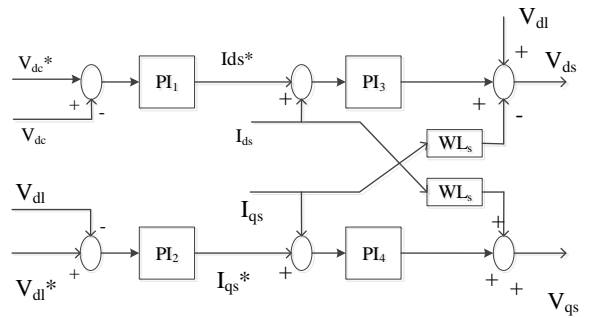

Fig. 3. Schematic configuration of the double-loop control strategy

This method has two loops; the external loop of active and reactive currents provides a reference current for maintaining the voltage at the PCC point, and the internal loop is used to control the inverter current [28]. PI controller coefficients have constant values that may not have an acceptable function in systems with time and nonlinear delay. This control strategy also requires four PI controllers for its control system, so it can be described as a tedious and time-consuming task for engineers to carry out trial and error studies to find suitable parameters when the operating conditions of the system have changed considerably [29].

*A. The DOV control strategy*

Relying on the principle of power balance, the transient output power of STATCOM is equal to the total power consumption of $R_s$, $L_s$ and load power $P_L$ and $Q_L$ [1].

$$P_S = P_{RL} + P_L$$
$$Q_S = Q_{RL} + Q_L \quad (3)$$

The STATCOM transient output power can also be expressed as follows:

$$P_s = \frac{3}{2}(V_{ds}I_{ds} + V_{qs}I_{qs})$$
$$Q_s = \frac{3}{2}(V_{ds}I_{qs} - V_{qs}I_{ds}) \quad (4)$$

The transient power consumption of $R_s$ and $L_s$ are expressed as follows:

$$P_{rl} = \frac{3}{2}R_s I^2 = \frac{3}{2}R_s(I_{ds}^2 + I_{qs}^2)$$
$$Q_{rl} = \frac{3}{2}wL_s I^2 = \frac{3}{2}wL_s(I_{ds}^2 + I_{qs}^2) \quad (5)$$

And the active and reactive power equations are defined as follows:

$$P_l = \frac{3}{2}(V_{dl}I_{ds} + V_{ql}I_{qs})$$
$$Q_l = \frac{3}{2}(V_{dl}I_{qs} - V_{ql}I_{ds}) \quad (6)$$

These equations are transmitted using the park transformation to the d-q axis, on which the voltage of the PCC point corresponds to that of the d-axis, and the q-axis is perpendicular to the d-axis:

$$V_{dl} = |v|$$
$$V_{ql} = 0 \quad (7)$$

By inserting Eq. (7) in Eq. (6), the following equation is obtained:

$$P_l = \frac{3}{2}(V_{dl}I_{ds})$$
$$Q_l = \frac{3}{2}(V_{dl}I_{qs}) \quad (8)$$

By inserting Eqs. (4), (5) and (8) in Eq. (3), the following equation is obtained:

$$V_{ds} = R_s I_{ds} - wL_s I_{qs} + V_{dl}$$
$$V_{qs} = R_s I_{qs} + wL_s I_{ds} \quad (9)$$

Therefore, the STATCOM reference output voltage ($V_{ds}$, $V_{qs}$) can be obtained using the STATCOM reference current control ($I_{ds}$, $I_{qs}$), $R_s$, $L_s$ and $V_{dL}$ as shown in Figure 4.

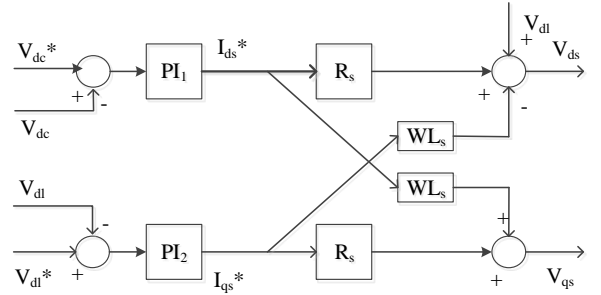

Fig. 4. Schematic configuration of the DOV control strategy[1]

*B. The proposed control strategy*

In this strategy, the power equations and the STATCOM are first defined and described, and then, using the obtained equations, a new control system is designed.

The active and reactive power equations in a permanent state are defined as follows:

$$P_l = \frac{3}{2}(V_{dl}I_{ds} + V_{ql}I_{qs})$$
$$Q_l = \frac{3}{2}(V_{dl}I_{qs} - V_{ql}I_{ds}) \quad (10)$$

These equations are transmitted using Park's transformation to the d-q axis, on which the voltage of the PCC point corresponds to that of the d-axis, and the q-axis is perpendicular to the d-axis:

$$V_{dl} = |v|$$
$$V_{ql} = 0 \quad (11)$$

And by inserting Eq. (11) in Eq. (10), the following equation is obtained:

$$P_l = \frac{3}{2}(V_{dl}I_{ds})$$
$$Q_l = \frac{3}{2}(V_{dl}I_{qs}) \quad (12)$$

As shown in Eq. (12), in this control method, there is an active power control as well as a reactive power control. The active power can be controlled using the d-axis current component, and the reactive power using the q-axis current component.

By inserting Eq. (11) in Eq. (2), the following equation is obtained:

$$L_s \frac{dI_{bs}}{dt} = -R_s I_{ds} + wL_s I_{qs} + V_{ds} - V_{dl}$$
$$L_s \frac{dI_{qs}}{dt} = -R_s I_{qs} - wL_s I_{ds} + V_{qs} \tag{13}$$

Eq. (13) can be expressed as follows:

$$V_{ds} = R_s I_{ds} - wL_s I_{qs} + L_s \frac{dI_{bs}}{dt} + V_{dl}$$
$$V_{qs} = R_s I_{qs} + wL_s I_{ds} + L_s \frac{dI_{qs}}{dt} \tag{14}$$

Therefore, the STATCOM output voltage ($V_{ds}$, $V_{qs}$) can be obtained by controlling the STATCOM current control ($I_{ds}$, $I_{qs}$), $R_s$, $L_s$ and $V_{dL}$ as shown in Figure 5.

This system includes only two PIs that are used to regulate the AC voltage and DC voltage. In the AC voltage regulator, The input is the difference between the terminal voltage and reference voltage and the output is a q-axis current. The DC voltage regulator receives the DC voltage error as an input and generates d-axis current. since the d-axis is always coincident with the voltage vector of the PCC, so it is an effective factor in passing the active power; also the q-axis is in quadrature with the voltage, so it is an effective factor in passing the reactive power through the converter .

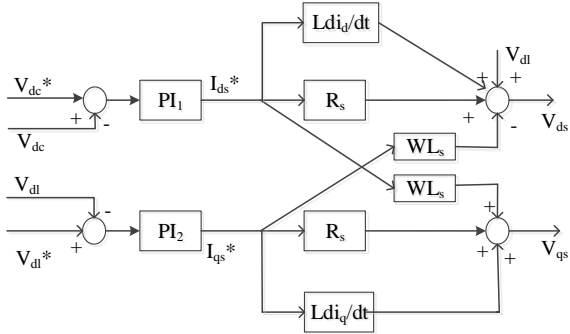

Fig. 5.  The proposed controller of

The active power will change the voltage of the DC bus and the reactive power will have the same effect on the terminal voltage. Then, using $I_q$ and $I_d$, the reference value of the STATCOM output voltage is obtained through Eq. (14). As can be seen in Fig. 5, this strategy has fewer PIs than the double-loop controller.

## IV. SIMULATION RESULTS

In this section, the transient performance of the induction motors is simulated in the presence of a STATCOM in Matlab/Simulink. The studied system consists of 9 induction motors connected through a transformer, and a transmission line to a three-phase network. As shown in Fig. 6, in parallel with the electric load, a three-phase reactive power compensator is used to improve the dynamic behavior of the system.

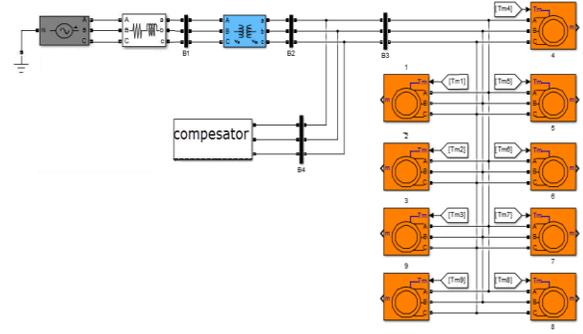

Fig. 6.  The studied circuit system with compensation in Simulink

For accuracy and speed evaluation, the simulation results are compared with DOV and double-loop controllers. A 20-percent voltage swell is applied on load bus for duration of 2 seconds; then, at t = 10s, a 20-percent voltage sag is applied for duration of 2 seconds.

Fig. 7 shows the system terminal voltage variation per unit for different controllers. when using a conventional controller. As can be seen from Fig. 7, proposed controller yields better performance compared to the other controllers. The propose controller has better compensation in comparison with double-loop controller and less overshoot in comparison with DOV method.

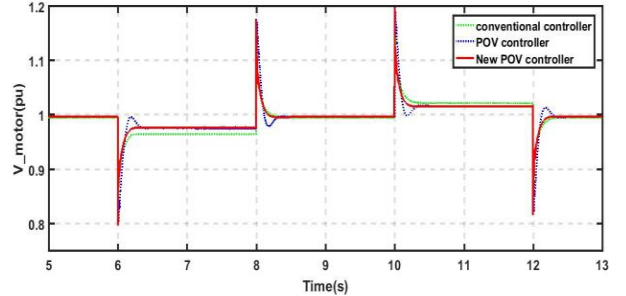

Fig. 7.  Terminal voltage of motor

Fig. 8 shows the changes in the speed of the induction motors. By compensating the voltage drop of the motors, the proposed controller reduces the speed variations in comparison with double-loop controller and less overshoots than with the DOV controller.

Fig. 9 shows the torque of the induction motors. As can be seen from the figure, the proposed controller yields better performance compared to the other controllers.

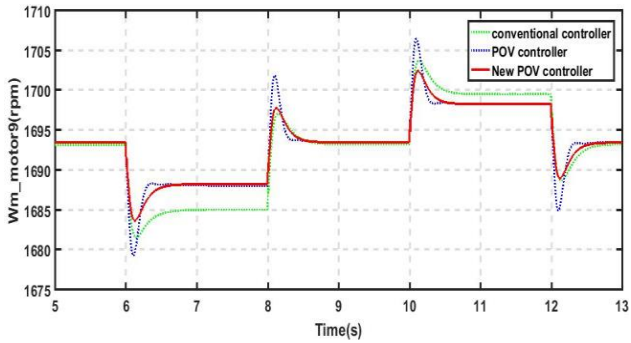

Fig. 8. Rotor speed (motor 3)

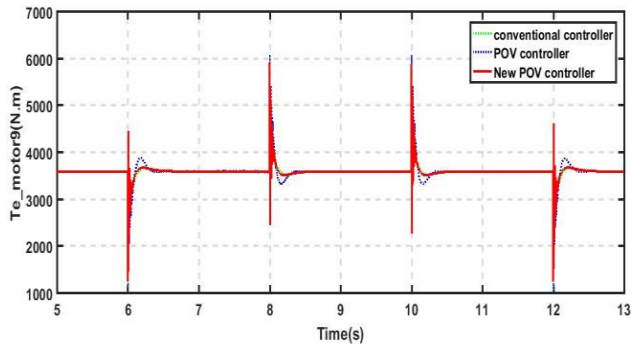

Fig. 9. the torque of the induction motors

V. CONCLUSION

This study shows that there are a number of problems with conventional STATCOM controllers, such as the large number of PI controllers and the difficulties posed when regulating their parameters. DOV controllers, too, present their own set of problems, including frequent overshoots and a slower dynamic response. In this paper, an improved DOV control strategy is presented for use with a STATCOM to overcome these problems. To investigate the proposed control function, the system was tested under a voltage sag and a voltage swell, where the controller was used to reduce overshoots and increase response speed. A comparison of the simulation results of the proposed controller to those of a DOV controller and a conventional controller indicated that this controller performs better under voltage swells and sags and reaches its final value with a higher speed after correcting the fault. This controller has fewer PIs and overshoots than with the conventional controller.